\documentclass[preprint,12pt,3p]{elsarticle}
\journal{Economic Analysis and Policy}

\usepackage{matlab-prettifier}
\usepackage{lineno}
\usepackage{graphicx}
\usepackage{amsmath}
\usepackage{xfrac}
\usepackage{parskip}
\usepackage[backref=page]{hyperref}
\usepackage{adjustbox}
\usepackage{setspace}
\hypersetup{ 
backref=true, 
bookmarks=true, 
unicode=false, 
pdftoolbar=true, 
pdfmenubar=true, 
pdffitwindow=false, 
pdfstartview={FitH}, 
pdftitle={}, 
pdfauthor={}, 
pdfsubject={}, 
pdfkeywords={}{}, 
pdfnewwindow=true, 
colorlinks=true, 
linkcolor=BlueViolet, 
citecolor=maroon, 
urlcolor=blue,
}
\usepackage{comment}
\usepackage{subfig}
\usepackage{caption}
\usepackage{longtable}
\usepackage{amssymb}
\usepackage{gensymb}
\usepackage{siunitx}

\usepackage{natbib}
\biboptions{comma,round}
\setcitestyle{authoryear}

\begin{document}
\begin{frontmatter}

\title{The fiscal implications of stringent climate policy}

\author[label1,label2,label3,label4,label5,label6,label7]{Richard S.J. Tol\corref{cor1}\fnref{label8}}
\address[label1]{Department of Economics, University of Sussex, Falmer, UK}
\address[label2]{Institute for Environmental Studies, Vrije Universiteit, Amsterdam, The Netherlands}
\address[label3]{Department of Spatial Economics, Vrije Universiteit, Amsterdam, The Netherlands}
\address[label4]{Tinbergen Institute, Amsterdam, The Netherlands}
\address[label5]{CESifo, Munich, Germany}
\address[label6]{Payne Institute for Public Policy, Colorado School of Mines, Golden, CO, USA}
\address[label7]{College of Business, Abu Dhabi University, UAE}

\cortext[cor1]{Jubilee Building, BN1 9SL, UK}
\fntext[labe8]{This paper was presented at 7th Shanghai-Edinburgh-London Green Finance Conference, the 6th Ethical Finance and Sustainability Conference, and the Commodity and Energy Markets Association Annual Conference 2023.}

\ead{r.tol@sussex.ac.uk}
\ead[url]{http://www.ae-info.org/ae/Member/Tol\_Richard}

\begin{abstract}
Stringent climate policy compatible with the targets of the 2015 Paris Agreement would pose a substantial fiscal challenge. Reducing carbon dioxide emissions by 95\% or more by 2050 would raise 7\% (1-17\%) of GDP in carbon tax revenue, half of current, global tax revenue. Revenues are relatively larger in poorer regions. Subsidies for carbon dioxide sequestration would amount to 6.6\% (0.3-7.1\%) of GDP. These numbers are conservative as they were estimated using models that assume first-best climate policy implementation and ignore the costs of raising revenue. The fiscal challenge rapidly shrinks if emission targets are relaxed. 
\end{abstract}

\end{frontmatter}

\newpage

\section{Introduction}
Much has been written about how to reduce greenhouse gas emissions and how much that would cost \citep[see][for a review of recent studies]{Riahi2022IPCC} but there is little about the implications for the public finances. This is an odd omission. Rapid emission reduction requires a major overhaul of the energy sector and energy-intensive activities \citep{IEA2021}. The energy transition will not just affect energy but everything it touches, including tax revenue and government spending. \citet{IEA2022}, for instance, reports that investment in the energy sector needs to double between 2020 and 2030, from 2\% to 4\% of GDP. This paper uses results from commonly-used integrated assessment models to study the impact of stringent climate policy on tax revenue and public expenditure, revealing the potential size of the carbon industry in the process.

The climate economics literature has focused on how best to reduce emissions \citep{Dubash2022IPCC} and what that would cost \citep{Riahi2022IPCC}. Much attention has been paid to the technical feasibility of rapid emission reduction \citep{Clarke2009} and to the required transition of energy, agriculture, and transport. The accompanying changes in the public sector have been largely ignored, with one exception, namely how to best use the revenues from a carbon tax (or permit auction). Using such revenue to reduce other taxes, which hold back economic growth or job creation, could result in a double dividend \citep{Goulder1995} or, if the income distribution improves too, a triple dividend \citep{vanHeerden2006}; \citet{Distefano2023} explore a quadruple dividend, adding public debt to the mix.

The multiple-dividend literature is focused on the structure of tax revenue, but ignores its size. Indeed, for analytical clarity, these papers \emph{assume} budget-neutrality. \citet{Belfiori2023} show that revamped consumption, energy, and income taxes can be a first-best policy, correcting the climate externality without an explicit Pigou tax. However, \citet{Tol2012CCL} argues that stringent climate policy may well require an overall increase in tax revenue and so lead to an expansion of the state. 

\begin{figure}[p]
    \centering
    \includegraphics[width=\textwidth]{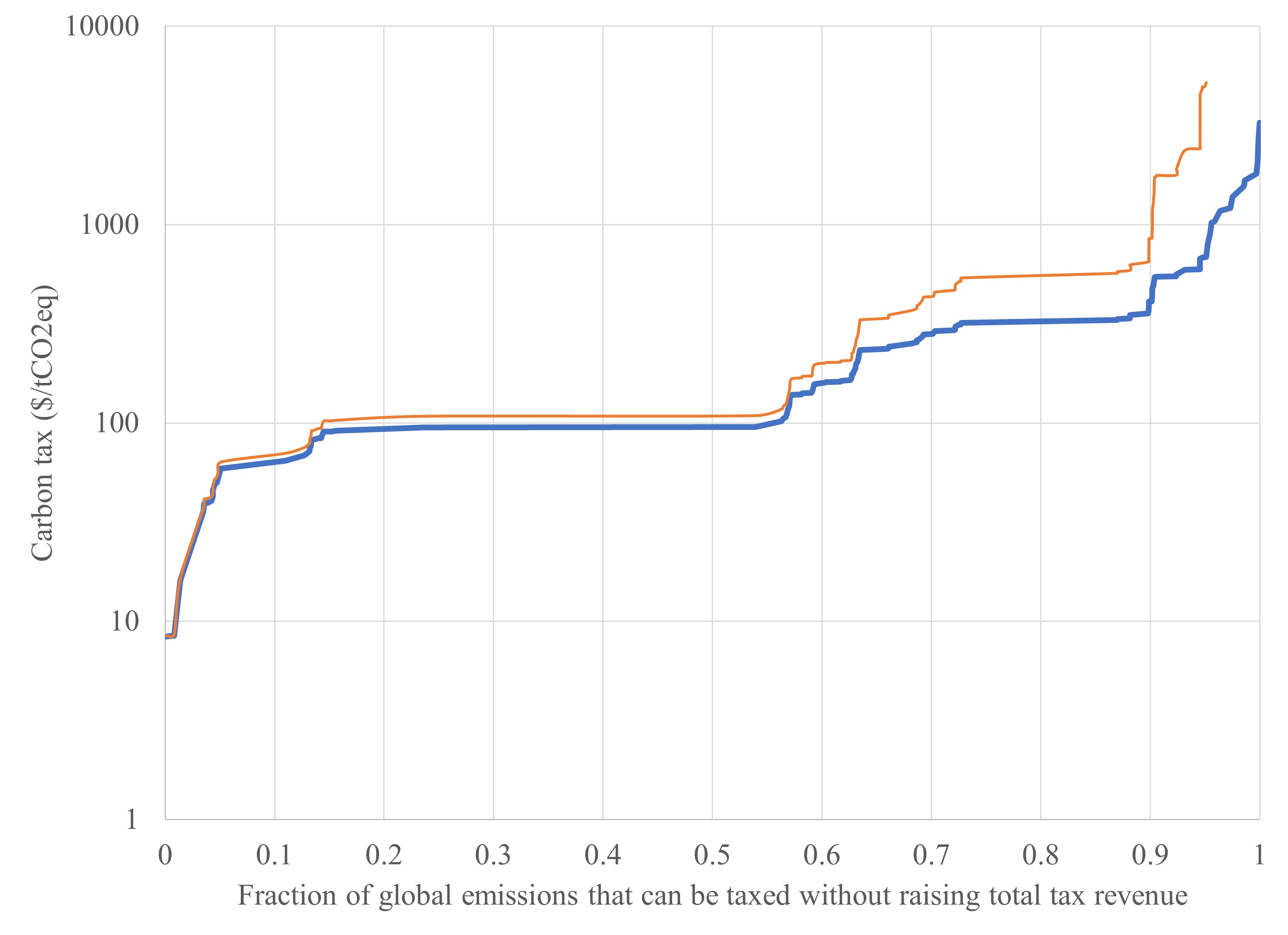}
    \caption{The Leviathan tax for 2019 without (thick blue line) and with emission reduction (thin orange line}
    \label{fig:leviathan}
\end{figure}

\citet{Tol2012CCL} defines the \emph{Leviathan tax} as that carbon tax whose revenue could replace the revenue of all other taxes combined.\footnote{Note that no assumptions are made on the desirable level of total tax revenue.} Figure \ref{fig:leviathan} shows the Leviathan tax for 2019. It is calculated as the greenhouse gas emission intensity of the economy\textemdash emissions over output\textemdash times the tax revenue as a share of GDP. Data are available from the World Bank for 145 countries. Figure \ref{fig:leviathan} ranks these countries by their Leviathan tax, and plots this against their share in global emissions. The Central African Republic has the lowest Leviathan tax: A carbon tax of \$8/tCO\textsubscript{2eq} would be budget-neutral if all other taxes are abolished. Sweden has the highest Leviathan tax: \$3,263/tCO\textsubscript{2eq}. The global average is \$242/tCO\textsubscript{2eq}

The Sixth Assessment Report of Working Group III of the Intergovernmental Panel on Climate Change \citep{Riahi2022IPCC} reports that, according to the median model, a carbon tax of around \$100/tCO\textsubscript{2eq} is needed in 2030 to have a good chance of meeting the 2\celsius target of the 2015 Paris Agreement. India's Leviathan tax (for 2019) is \$95/tCO\textsubscript{2eq}, China's \$96/tCO\textsubscript{2eq}, and Indonesia's \$102/tCO\textsubscript{2eq}. Stringent climate policy is therefore not just a technical and economic challenge, but a fiscal challenge too.

Fiscal problems would arise long before the Leviathan tax is reached. \citet{Besley2013} show that fiscal capacity has grown slowly and that the structure of tax revenues has developed gradually. Rapid, massive change in tax collection is unprecedented and would be difficult, or so the historical record suggests. Climate policy would require two tax revolutions. First, taxes should shift to carbon from everything else to drive emissions to zero\textemdash and then taxes would have to shift back to maintain tax revenue.

\citet{Dowlatabadi2000} was perhaps the first to warn about possible tax revolts \citep{Burg2004, Keen2021} in the context of climate policy. One example is the 2018 protests by \textit{les gilets jeunes} in France in response to a modest carbon tax on transport fuels \citep{STOLL2021}. The carbon taxes needed to meet the Paris targets are not modest\textemdash and they will need to apply in countries that are not as used to high taxes as France is.

Throughout the paper, I write about climate policy as if a carbon tax were the sole policy instrument. The reason for this is that the models I rely on make this assumption. Although the optimal climate policy is a carbon tax, a uniform carbon tax, and nothing but a carbon tax \citep{Tol2023bk}, the bulk of past and present climate policies rely on other instruments. There is no reason to assume future climate policy will be any different.

Some of the insights carry over. Cap-and-trade with auctioned permits behaves much like a carbon tax, the key difference being that permit prices fluctuate and taxes do not. The revenue of permit auctions can be used to reduce taxes.

If permits are grandparented instead of auctioned, climate policy is like a carbon tax (at the margin) plus lump-sum capital subsidies for the recipients of free permits. These capital subsidies pose no burden on the fiscal budget as the government costlessly creates the permits before giving them away. In this case, taxes cannot be reduced. Instead, the public sector expands.

Subsidies, another popular policy instrument, are negative taxes. Other taxes would need to go up substantially if subsidies are used to reduce emissions at the required scale.

Any technical standard has an equivalent tax \citep{Baumol1971}. If standards are the policy instrument of choice\textemdash as they often are\textemdash the tax burden calculated below is a measure of the changes needed in the economy. Fiscal implications would be indirect.

More troublesome than the assumption of a carbon tax is the assumption, again taken from the models I rely on, that climate policy will be cost-effective.\footnote{This paper shies away from a discussion of optimal climate policy targets, which are treated extensively elsewhere \citep{Nordhaus1992, Tol1999kyoto, Tol2012EP, Tol2023NCC}.} Current climate policy most definitely is not \citep[e.g.,][]{Grimm2022}. However, this strengthens the argument below. If \emph{cost-effective} policy implies unrealistically large fiscal shocks, then \emph{sub-optimal} policy (with the same emissions target) implies even larger shocks. Admittedly, without a carbon tax, those shocks may not be to the public finances; they will be to the economy instead.

The paper proceeds as follows. Section \ref{sc:methods} discusses the materials and methods used. Section \ref{sc:results} presents the results. Section \ref{sc:conclude} concludes.

\section{Materials and methods}
\label{sc:methods}
The \href{https://data.ece.iiasa.ac.at/ar6/}{IPCC AR6 scenario database} contains projections of GDP, greenhouse gas emissions, carbon dioxide sequestration, and emission taxes for a range of \textit{ex-ante} models and a range of scenarios with and without emission reduction targets. The database contains a host of variables on the structure of energy demand and supply, agriculture, land use, and so on. I here only use GDP, gross carbon dioxide emissions, gross carbon uptake, and carbon taxes. For most models, results are reported for 10-year intervals until 2100.

While generally well-structured, the database, unfortunately, does not match baseline and policy scenarios; this was added, manually, based on scenario names. Missing rows were replaced by missing observations. This then leads to the percentage reduction of GDP and emissions from baseline.

Total carbon tax revenue (subsidy) follows from multiplying gross carbon dioxide emissions (sequestration) with carbon taxes. 

As highlighted by \citet{Riahi2022IPCC}, the models in the database show a wide range of results. This is not a surprise, as the models have different structures and use different assumptions on economic growth, on relative prices, on technological change, on income, price and substitution elasticities, and on reserves, resources and potentials. Some models are computable general equilibrium models, others energy system models, and yet others are growth, econometric or new Keynesian models. All models have some foresight, many perfect foresight. The only commonality is that all models have been used to study \emph{future} climate policy.

Note that I do not correct the IPCC database for reporting bias \citep{Tavoni2010}. This omission likely leads to an underestimate of the true cost of climate policy.

I follow \citet{Tol2014en} and compare these \textit{ex-ante} models to the data, but where \citet{Tol2014en} relied on a fairly basic statistical analysis, I here use five advanced econometric studies of the efficacy of carbon pricing \citep{Rafaty2020, Kohlscheen2021, Sen2018, Metcalf2020, Best2020}. These \textit{ex-post} studies use different estimators and different samples, but they all study the effect of \emph{past} climate policy on past emissions. The efficacy of a carbon tax is here defined as the percentage emission reduction per dollar per tonne of carbon dioxide carbon tax. This measure is reported by, or easily derived from the five econometric studies. It is also readily calculated from the data in the IPCC AR6 scenario database.

I use Bayesian statistics to assess the credibility of the different models. I use a non-informative prior. The results of the econometric models are the likelihood. Combined, this gives the posterior estimate of the tax efficacy. Alternatively, I shrunk the five estimates to a single, combined one \citep{Goldberger1964}. In a second step, as a prior, I assumed that each IPCC model is equally likely. The posterior likelihood of the tax efficacy implies a probability that an \textit{ex-ante} model is able to reproduce observed climate policy as measured by the \textit{ex-post} models. 

While the methods are well-established, this is their first application to the fiscal implications of stringent climate policy.

\section{Results}
\label{sc:results}

\subsection{Model skill}
Before discussing the key results, I need to establish which model is most credible. This is because the range of model range is so large. Some models find that climate policy is too cheap to meter, others that it would lead to economic ruin.

Table \ref{tab:taxefficacy} shows the efficacy of a carbon price for the 24 models in the \href{https://data.ece.iiasa.ac.at/ar6/}{IPCC AR6 scenario database} for which this information was available. Tax efficacy is the percentage CO\textsubscript{2} emission reduction (from baseline) in 2030 divided by the carbon tax or permit price in the same year. (Recall that the models assume foresight.) Efficacy differs by \emph{three} orders of magnitude from 0.0042\%/\$ for \textsc{ices} to 4.8\%/\$ for \textsc{coffee}.

At the bottom of \ref{tab:taxefficacy}, five econometric estimates of the same metric are shown \citep{Rafaty2020, Kohlscheen2021, Sen2018, Metcalf2020, Best2020}. Three of these studies agree that a carbon price of \$1/tCO\textsubscript{2} would cut emissions by some 0.1\%, higher than 2 of the 24 IPCC models and lower than 21. The other two econometric studies find that carbon pricing is more effective. The minimum and maximum differ by one order of magnitude.

The posterior mean, weighted average, or shrunk estimate is a reduction of 0.13\% per dollar per metric tonne of carbon dioxide. This implies, assuming linearity, that a carbon tax of \$792/tCO\textsubscript{2} would fully decarbonize the world economy.

The short-run Leviathan tax is discussed in the introduction. It assumes that the imposed carbon tax does not affect emissions. Figure \ref{fig:leviathan} also shows the long-run Leviation tax, using the central estimate of 0.13\% emission reduction per dollar carbon tax. The Leviathan tax increases, but not sufficiently so that the IPCC's \$100/tCO\textsubscript{2} carbon tax looks materially less problematic. 

Only the \textsc{imaclim} model \citep{Crassous2006, Sassi2010, Waisman2012, Bibas2015, Mejean2019} is close to the majority of the empirical evidence.\footnote{I have criticized this model for having so many distortions that it is hard to interpret the results. That said, the economy is full of distortions.} Indeed, 95.5\% of the posterior probability mass goes to \textsc{imaclim}. The posterior probability of \textsc{gemini} is 0.5\%. The probabilities of the remaining models are very small.

\subsection{The impact of stringent climate policy}
Table \ref{tab:taxes} shows the main result. Twelve models in the IPCC AR6 database report scenarios that cut global carbon dioxide emissions by 95\% or more in 2050. Table \ref{tab:taxes} shows the carbon price and the value of carbon capture and emissions, all averaged across the scenarios for each of the models. The carbon price is either the explicit carbon tax, the price of tradable permits, or the shadow price of the emissions constraint. The value of emissions is the total revenue of either a carbon tax or the auction of carbon permits. The value of carbon capture is either the total expenditure on carbon removal subsidies or the sum total spent on carbon offsets. Both values are given as a share of GDP.

The results vary widely. The most optimistic model is again the \textsc{coffee} model. As in Table \ref{tab:taxefficacy}, this model finds that a minimal carbon tax would completely decarbonize the economy. Revenues and expenditures are therefore small too. At the other extreme, \textsc{dne21} has a carbon tax revenue of 3 times GDP, and on top spends 2 times GDP on carbon removal. One would hope this is a reporting error rather than a genuine result of what would be a mistaken model.

Discarding the two outliers, carbon tax revenue ranges from 1 to 17\% of GDP. This range is wide. A tax reform that brings in 1\% of GDP by 2050 is feasible. Tax reforms at this scale happen regularly \citep{OWID2016}. The high end of the range is more difficult. The global average tax revenue was 14\% of GDP in 2019.\footnote{See \href{https://data.worldbank.org/indicator/GC.TAX.TOTL.GD.ZS?view=chart}{World Bank}.} An expansion of the public sector by 3\% in 30 years is doable. Reducing if not abolishing all other taxes would, of course, be an election winner\textemdash although taxes are rarely abolished \citep{Seelkopf2021}. However, as emissions approach zero, the tax base would get narrower and narrower and the carbon tax higher and higher, so that the fiscal system becomes increasingly distortionary. As emissions go to zero, so does carbon tax revenue\textemdash other taxes will have to be reintroduced, a politically more challenging prospect.\footnote{In \textsc{GCAM}, emissions fall to zero before 2050. Its fiscal transition is even faster.} \textsc{imaclim}, the most credible model, has total carbon tax revenues at 7\% of GDP in 2050, replacing ``only'' half of all other taxes (if government budget neutrality is assumed).

Total carbon removal subsidies, or payments for offsets, range from 0.3\% (\textsc{aim}) to 7\% (\textsc{grape}) of GDP. The model that compares best to the data, \textsc{imaclim}, is at the high end of this range. A subsidy that is a few tenths of a percent of GDP is no problem. Climate change has been a key concern of many people around the world for decades \citep{Leiserowitz2006, Lee2015, Rettig2023}\textemdash the vocal protests of a small minority notwithstanding. Spending a small fraction of income on solving the climate problem should not be a problem. However, expenditure is much larger at the high end of the range, roughly equal to expenditures on health care. Public spending on health care is like motherhood and apple pie\textemdash we all rely on doctors and nurses to heal ourselves and our loved ones, and we all have friends and family who work in medicine and who deserve a decent salary. Carbon capture is very different. It solves a distant and abstract problem, rather than one that is close and obvious like ill-health. If climate policy is successful, there is not much of a problem to solve anymore, making it harder to continue to justify spending large sums of money. In order to keep costs down, carbon capture will be done where land is cheap\textemdash that is, where few people live\textemdash and heavily mechanized. Paying 7\% of your income in taxes to keep grandma alive and your nurse friend in work is one thing. Paying 7\% to a multinational company to suck carbon dioxide out of the air in a faraway country is something else.

\subsection{Regional results}
\begin{figure}[p]
    \centering
    \includegraphics[width=\textwidth]{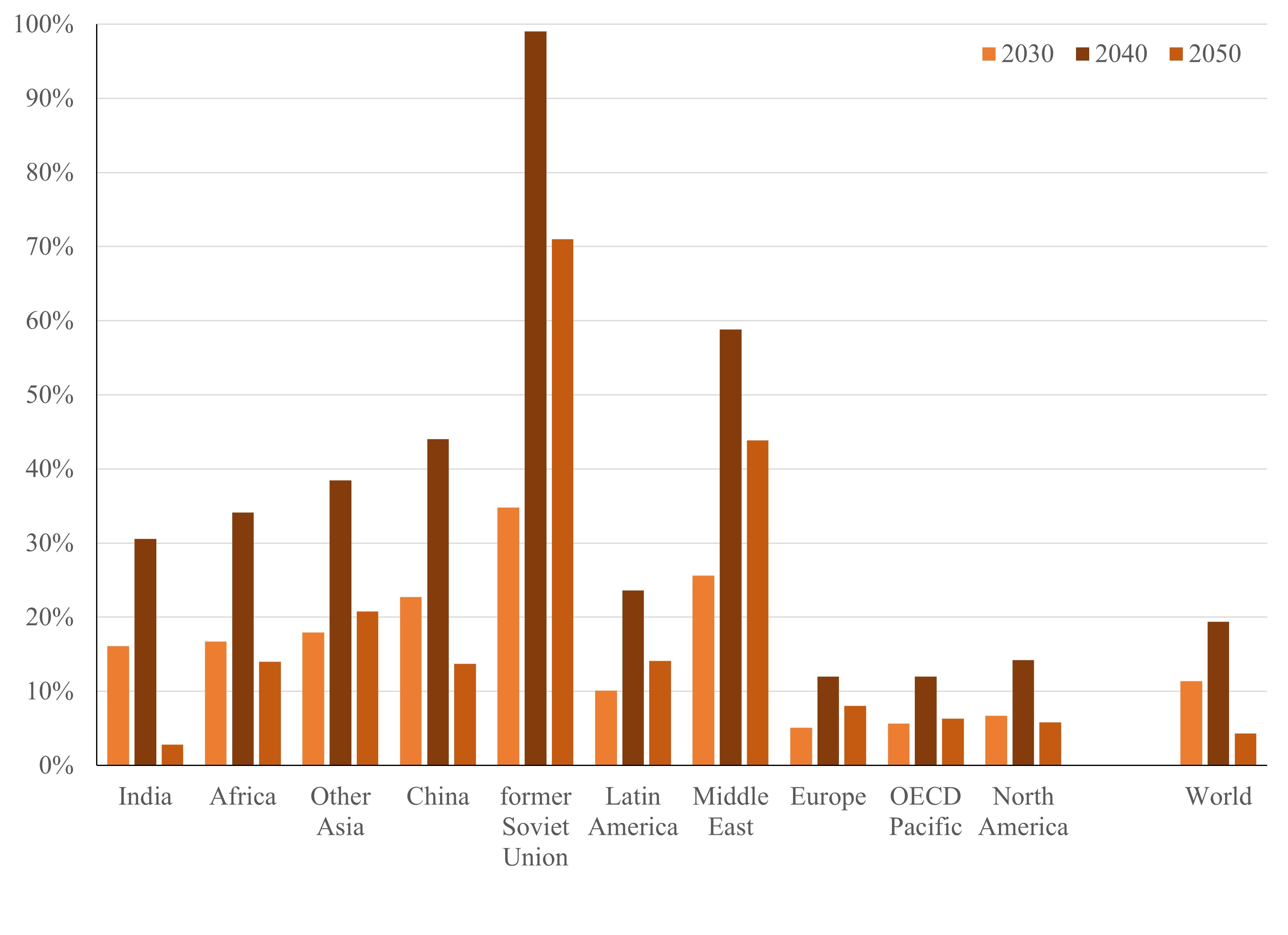}
    \includegraphics[width=\textwidth]{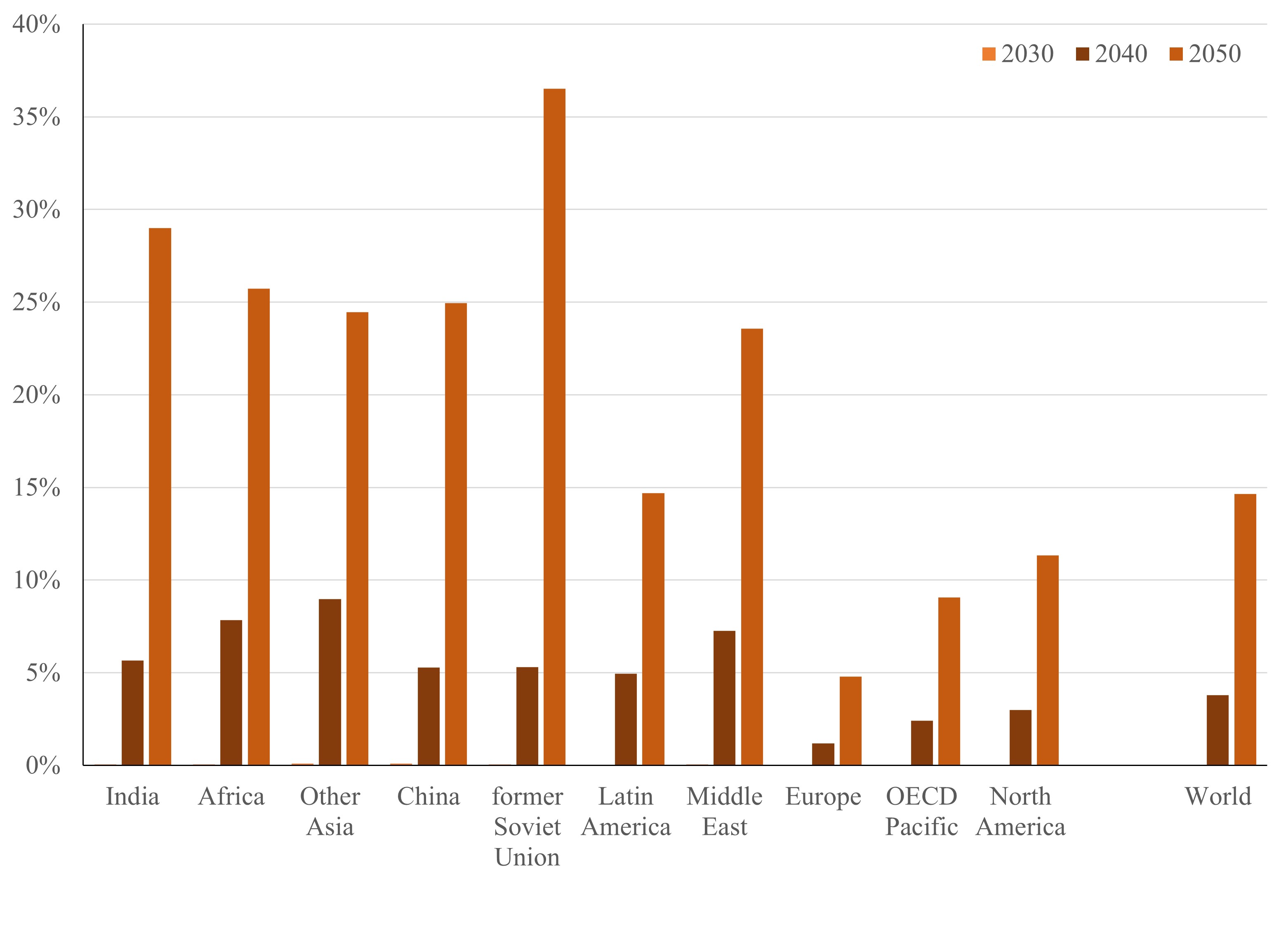}
    \caption{Carbon tax revenue (top panel) and carbon sequestration subsidy (bottom panel) as a share of GDP for 10 regions according to the \textsc{imaclim} model and its ADVANCE/2030/WB2C scenario.}
    \label{fig:regions}
\end{figure}

The above results are for the world as a whole. The models in the IPCC database also report regional results. I restrict the attention to \textsc{imaclim} and one particular scenario which reduces emissions by 94\% in 2050. The carbon tax is \$300/tCO\textsubscript{2} in 2030, rising to \$1,298/tCO\textsubscript{2} in 2040 and \$2,253/tCO\textsubscript{2} in 2050. Figure \ref{fig:regions} shows carbon tax revenue and sequestration subsidy, as a percentage of GDP, for 2030, 2040, and 2050.

Global carbon tax revenue is 4\% of GDP in 2050, a reasonable number, but 11\% in 2030 and 19\% in 2040\textemdash underlining yet again the fiscal challenge posed by stringent climate policy.

The results in Figure \ref{fig:regions} are ordered by per capita income in 2010. Carbon tax revenue is below the global average in the three richest regions, but above the global average in the seven poorest regions\textemdash with the exception of almost completely decarbonized India in 2050. The carbon tax revenue is very high in the carbon-intensive economies of the Middle East and the former Soviet Union.

The bottom panel of Figure \ref{fig:regions} shows the sequestration subsidies. The world total is 0.04\% of GDP in 2030, rising to 3.8\% in 2040 and 15\% in 2050. As with tax revenues, the numbers are lower for the three rich regions and higher for the seven poor regions. Note, however, that it may well be that there will substantial transfers between regions. This is less likely with direct subsidies, more likely with tradable permits and offsets.

That said, Figure \ref{fig:regions} highlights the scale of the activity. The sequestration sector would occupy almost 15\% of the world economy, over 35\% of the economy in the former Soviet Union.

\subsection{Results for more lenient climate policy}
The above results are for very stringent climate policy. Cutting carbon dioxide emissions by 95\% or more by 2050 is highly ambitious. The major fiscal implications highlighted above rapidly disappear for less stringent climate policy. This is because the fiscal implications are the product of carbon price and emissions. Take the subsidies for carbon dioxide removal first. A more lenient target would mean a lower volume at a lower price. The carbon tax revenue would fall too: Emissions would be higher but the carbon price lower; the former is linear, the latter exponential.

\begin{figure}[p]
    \centering
    \includegraphics[width=0.49\textwidth]{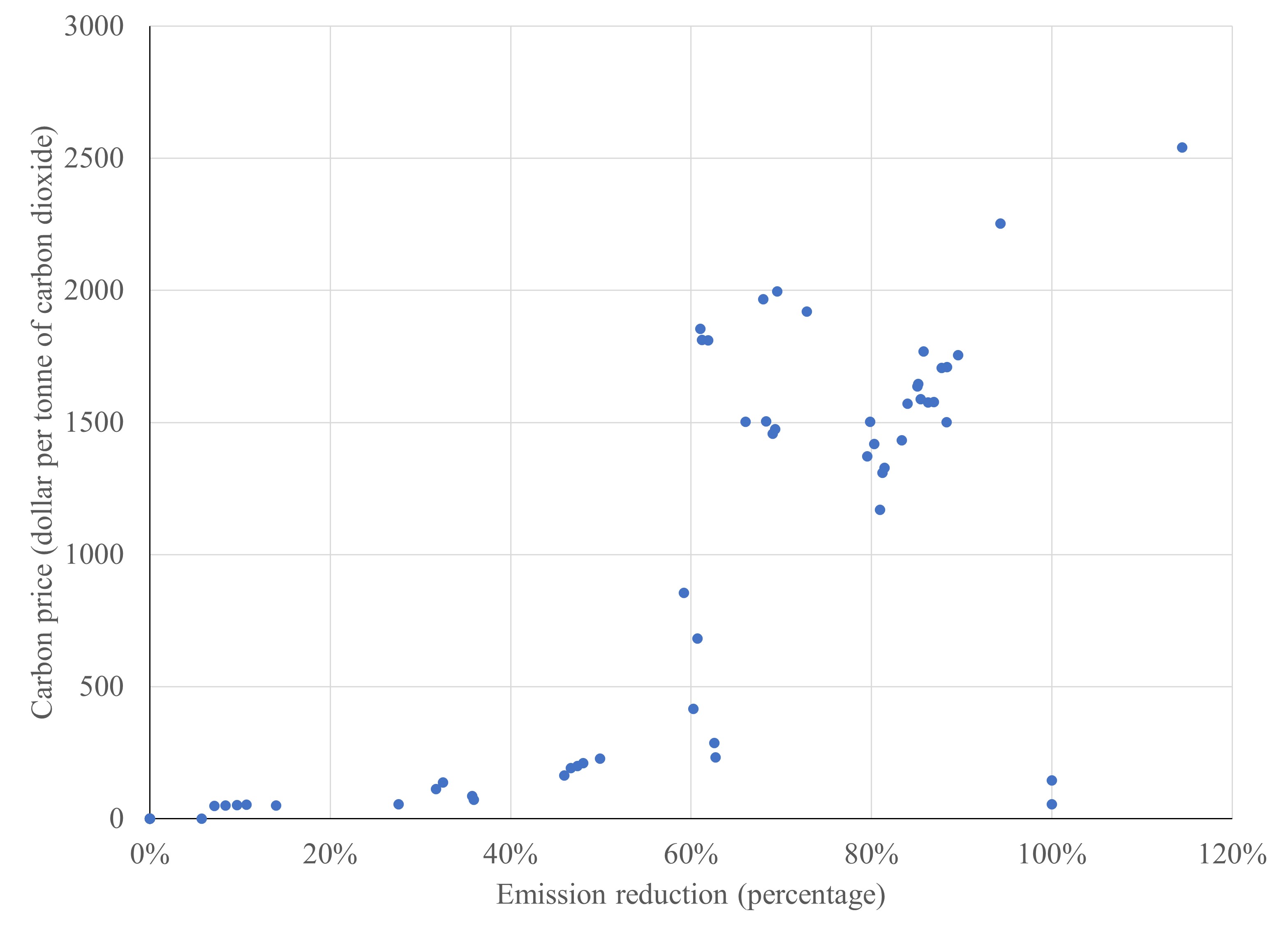}
    \includegraphics[width=0.49\textwidth]{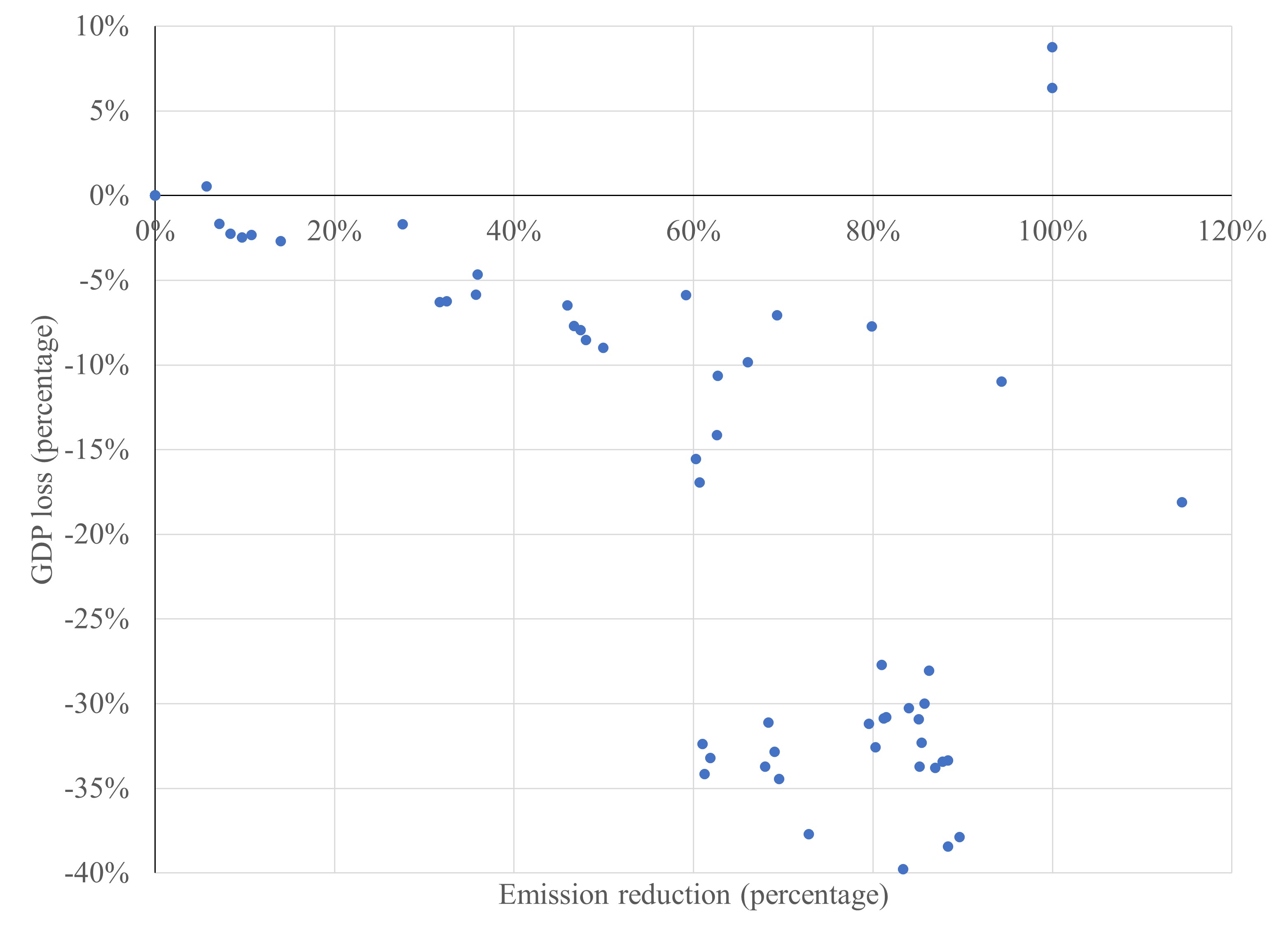}
    \includegraphics[width=0.49\textwidth]{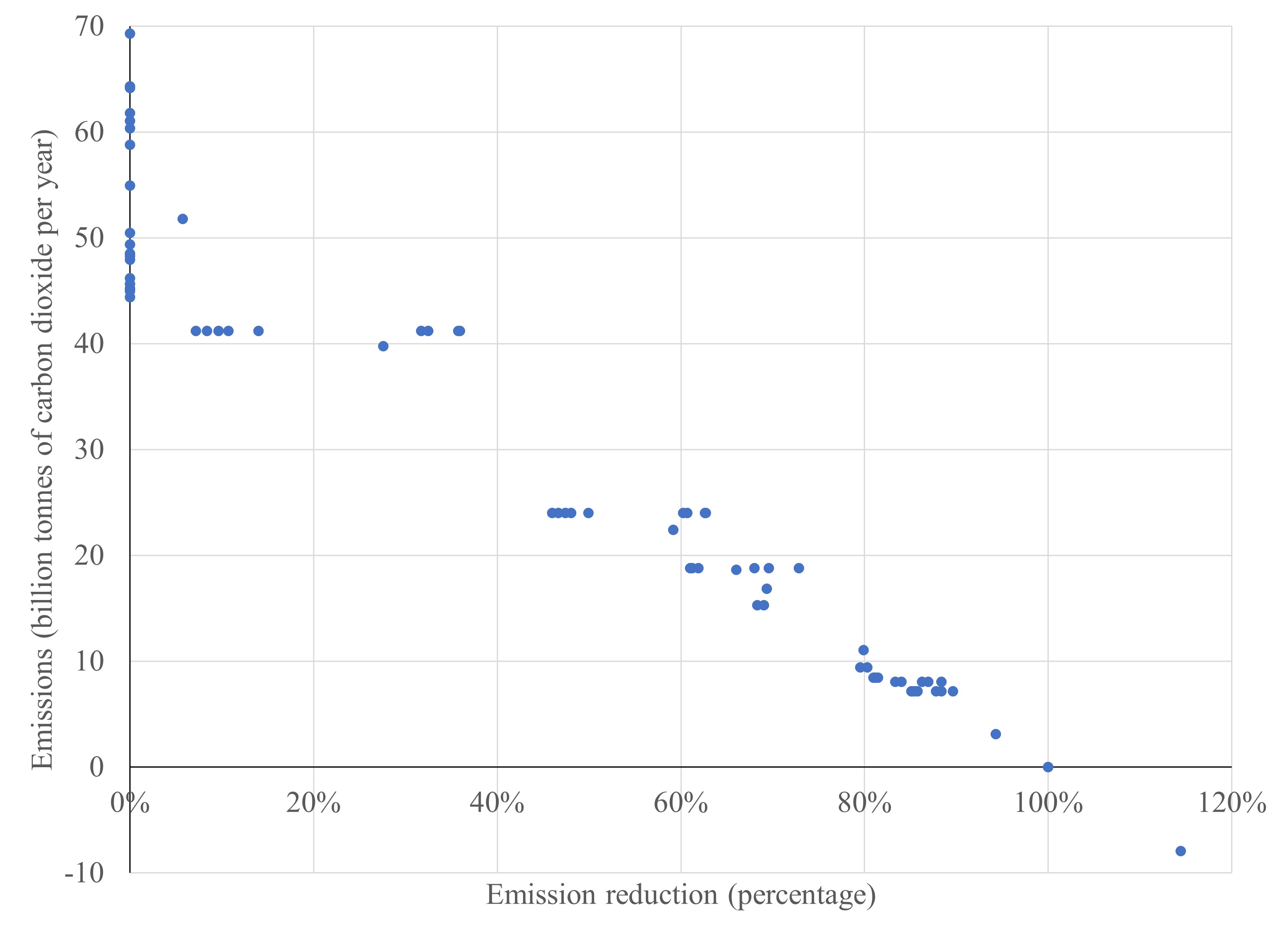}
    \includegraphics[width=0.49\textwidth]{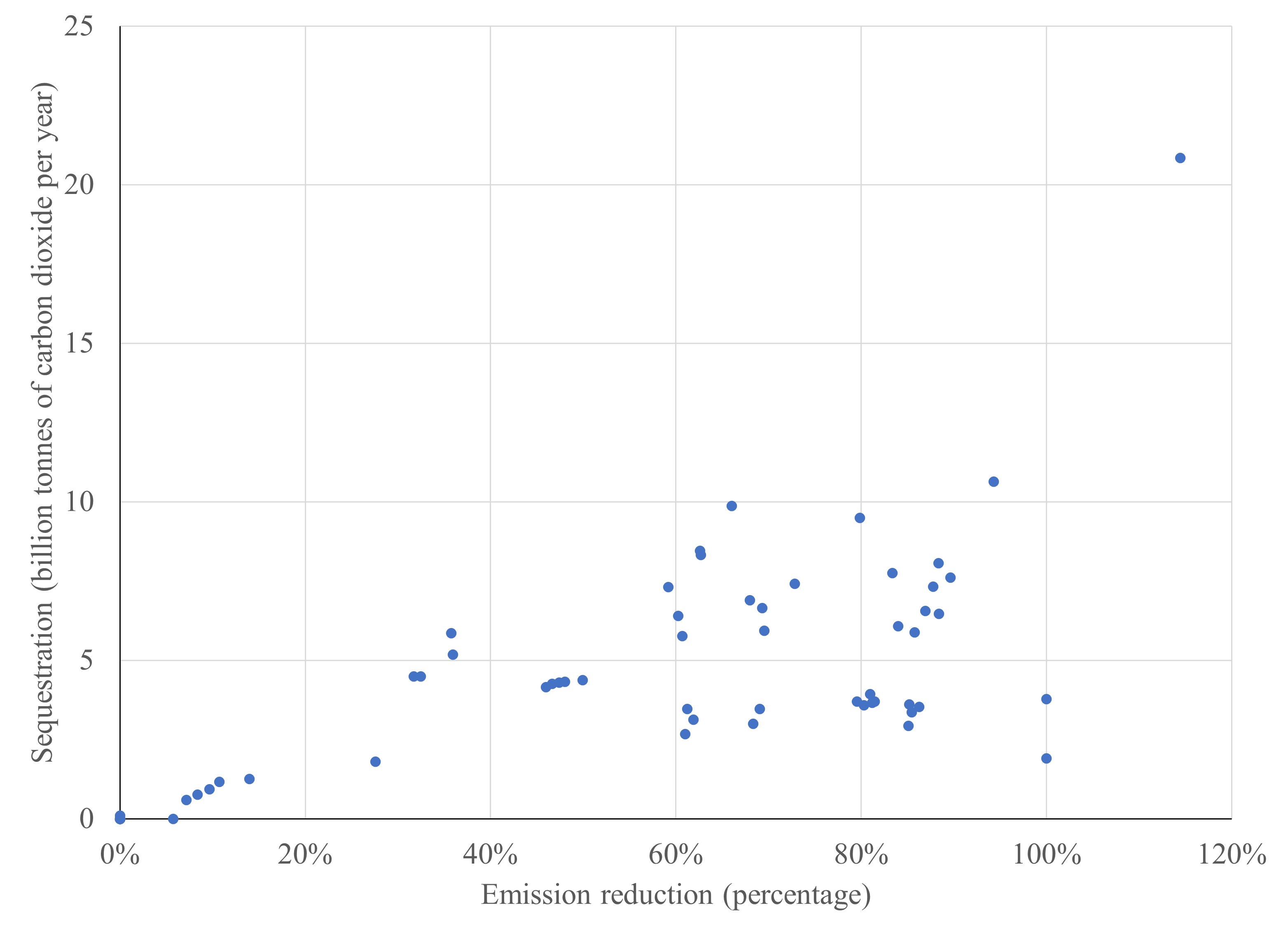}
    \caption{The carbon tax (top left), loss of GDP (top right), gross carbon dioxide emissions (bottom left) and carbon dioxide sequestration (bottom right) in 2050 as a function of carbon dioxide emission reduction from baseline according to the \textsc{imaclim} model.}
    \label{fig:imaclim}
\end{figure}

Figure \ref{fig:imaclim} illustrates this for the \textsc{imaclim} model for 2050. The top left panel plots the carbon price against emission reduction from baseline. The carbon price inches up until emissions are halved and then starts rising very quickly. However, emissions, shown in the bottom left panel, continue to fall steadily. Sequestration, in the bottom right panel, similarly shows no profound non-linearity. The top right panel shows the drop in GDP, which accelerates around a 50\% emission reduction. This accentuates carbon tax revenue and carbon sequestration expenditures relative to GDP.

\section{Discussion and conclusion}
\label{sc:conclude}
Stringent climate policy would pose a substantial fiscal challenge. The global revenue of the carbon tax needed to meet the targets of the 2015 Paris Agreement would be larger than the revenue of all other taxes combined, while a very large subsidy would need to be paid to remove carbon dioxide from the atmosphere. Tax revenues are larger still in poor parts of the world. Climate policy by other means than taxes and subsidies would shift, perhaps hide, probably exacerbate the fiscal burden.

The fiscal challenge rapidly shrinks as the emission reduction target becomes less stringent. The policy implication is thus to adopt a more lenient climate policy\textemdash or rather, as the gap between nominal targets and actual climate policy had widened \citep{UNEP2022}, to adopt more realistic rhetoric.

The implications for research are more profound. Model results show a very large range for the costs of future climate policy. This is partly inevitable. The future is inherently uncertain. However, the skill of \textit{ex-ante} models can be tested against over 30 years of experience with actual climate policy. This is here done with a single variable, tax efficacy, but these models generate many more variables, most of which are directly observed.

Even before testing their skills, two of the models in the IPCC database report patent nonsense. Either the database or the models need to be vetted better. The problems do not stop there. Many of the integrated assessment models used by the IPCC do not have a rich representation of the fiscal system\textemdash and none report this. Environmental regulation and general taxation interact \citep{Sandmo1975}. Ignoring prior tax distortions leads to unnecessarily expensive climate policy \citep{Barrage2019}. A tax is more distortionary as it rises and its base narrows\textemdash exactly what happens as emissions approach zero. Ignoring the excess burden in the climate policy endgame seems to be a crucial omission in integrated assessment models.

Unlike tax distortions, tax revolts are unpredictable\textemdash but the probability of tax revolts varies systematically with observable variables \citep{Dowlatabadi2000}. Dynamic stochastic general equilibrium models are now regularly used to study climate policy \citep{Cai2019, Bremer2021}. It strikes me that tax revolts are a key stochastic element. Tax revolts may be more likely if the costs of climate policy are distributed in a way that is seen to be unfair \citep{Chepeliev2021, Landis2021, Vandyck2021, Bohringer2022, Wu2022} and if assets are stranded and firms go bankrupt \citep{Davis2010, Tong2019, Ploeg2020jeem, Semieniuk2022, Flora2023}.

All this complicates climate policy and makes it more expensive. Adding the analytically convenient but unrealistic assumption of first-best policy implementation, it appears that policy-makers are ill-advised by the IPCC and its choice of models. More importantly, current emission reduction targets may need to be relaxed.

\bibliography{master}

\begin{thebibliography}{55}
\providecommand{\natexlab}[1]{#1}
\providecommand{\url}[1]{\texttt{#1}}
\expandafter\ifx\csname urlstyle\endcsname\relax
  \providecommand{\doi}[1]{doi: #1}\else
  \providecommand{\doi}{doi: \begingroup \urlstyle{rm}\Url}\fi

\bibitem[Barrage(2019)]{Barrage2019}
Lint Barrage.
\newblock {Optimal Dynamic Carbon Taxes in a Climate–Economy Model with
  Distortionary Fiscal Policy}.
\newblock \emph{The Review of Economic Studies}, 87\penalty0 (1):\penalty0
  1--39, 2019.
\newblock URL \url{https://doi.org/10.1093/restud/rdz055}.

\bibitem[Baumol and Oates(1971)]{Baumol1971}
W.~J. Baumol and W.~E. Oates.
\newblock The use of standards and prices for the protection of the
  environment.
\newblock \emph{Scandinavian Journal of Economics}, 73\penalty0 (1):\penalty0
  42--54, 1971.

\bibitem[Belfiori and Rezai(2023)]{Belfiori2023}
Elisa Belfiori and Armon Rezai.
\newblock Optimal climate policy: Making do with the taxes we have.
\newblock Technical report, Universidad Torcuato Di Tella, School of Business,
  Buenos Aires, 2023.

\bibitem[Besley and Persson(2013)]{Besley2013}
Timothy Besley and Torsten Persson.
\newblock Chapter 2 - taxation and development.
\newblock In Alan~J. Auerbach, Raj Chetty, Martin Feldstein, and Emmanuel Saez,
  editors, \emph{Handbook of Public Economics, vol. 5}, volume~5 of
  \emph{Handbook of Public Economics}, pages 51--110. Elsevier, 2013.
\newblock \doi{https://doi.org/10.1016/B978-0-444-53759-1.00002-9}.

\bibitem[Best et~al.(2021)Best, Hammerle, Mukhopadhaya, and Silber]{Best2021}
R.~Best, M.~Hammerle, P.~Mukhopadhaya, and J.~Silber.
\newblock Targeting household energy assistance.
\newblock \emph{Energy Economics}, 99, 2021.
\newblock \doi{10.1016/j.eneco.2021.105311}.

\bibitem[Best et~al.(2020)Best, Burke, and Jotzo]{Best2020}
Rohan Best, Paul~J. Burke, and Frank Jotzo.
\newblock {Carbon Pricing Efficacy: Cross-Country Evidence}.
\newblock \emph{Environmental \& Resource Economics}, 77\penalty0 (1):\penalty0
  69--94, 2020.
\newblock \doi{10.1007/s10640-020-00436-}.

\bibitem[Bibas et~al.(2015)Bibas, Méjean, and Hamdi-Cherif]{Bibas2015}
Ruben Bibas, Aurélie Méjean, and Meriem Hamdi-Cherif.
\newblock Energy efficiency policies and the timing of action: An assessment of
  climate mitigation costs.
\newblock \emph{Technological Forecasting and Social Change}, 90\penalty0
  (PA):\penalty0 137 – 152, 2015.
\newblock \doi{10.1016/j.techfore.2014.05.003}.

\bibitem[B\"{o}hringer et~al.(2022)B\"{o}hringer, Garc\'{i}a-Muros, and
  Gonz\'{a}lez-Eguino]{Bohringer2022}
Christoph B\"{o}hringer, Xaqu\'{i}n Garc\'{i}a-Muros, and Mikel
  Gonz\'{a}lez-Eguino.
\newblock Who bears the burden of greening electricity?
\newblock \emph{Energy Economics}, 105, 2022.
\newblock \doi{10.1016/j.eneco.2021.105705}.

\bibitem[Burg(2004)]{Burg2004}
David~F. Burg.
\newblock \emph{A World History of Tax Rebellions: An Encyclopedia of Tax
  Rebels, Revolts, and Riots from Antiquity to the Present}.
\newblock Routledge, {L}ondon and New York, 2004.

\bibitem[Cai and Lontzek(2019)]{Cai2019}
Y.~Cai and T.~S. Lontzek.
\newblock The social cost of carbon with economic and climate risks.
\newblock \emph{Journal of Political Economy}, 127\penalty0 (6):\penalty0
  2684--2734, 2019.
\newblock \doi{10.1086/701890}.

\bibitem[Chepeliev et~al.(2021)Chepeliev, Osorio-Rodarte, and {van der
  Mensbrugghe}]{Chepeliev2021}
Maksym Chepeliev, Israel Osorio-Rodarte, and Dominique {van der Mensbrugghe}.
\newblock Distributional impacts of carbon pricing policies under the {Paris
  Agreement}: {I}nter and intra-regional perspectives.
\newblock \emph{Energy Economics}, 102:\penalty0 105530, 2021.
\newblock URL
  \url{https://www.sciencedirect.com/science/article/pii/S0140988321004084}.

\bibitem[Clarke et~al.(2009)Clarke, Edmonds, Krey, Richels, Rose, and
  Tavoni]{Clarke2009}
Leon Clarke, Jae Edmonds, Volker Krey, Richard Richels, Steven Rose, and
  Massimo Tavoni.
\newblock International climate policy architectures: Overview of the {EMF 22}
  international scenarios.
\newblock \emph{Energy Economics}, 31\penalty0 (S2):\penalty0 S64--S81, 2009.
\newblock URL
  \url{http://www.sciencedirect.com/science/article/B6V7G-4XHVH34-2/2/67f06e207a515adba42f7455a99f648e}.

\bibitem[Crassous et~al.(2006)Crassous, Hourcade, and Sassi]{Crassous2006}
Renaud Crassous, Jean-Charles Hourcade, and Olivier Sassi.
\newblock Endogenous structural change and climate targets modeling experiments
  with {I}maclim-{R}.
\newblock \emph{Energy Journal}, 27\penalty0 (SPEC. ISS. MAR.):\penalty0 259
  – 276, 2006.

\bibitem[Davis et~al.(2010)Davis, Caldeira, and Matthews]{Davis2010}
Steven~J. Davis, Ken Caldeira, and H.~Damon Matthews.
\newblock Future {CO}\textsubscript{2} emissions and climate change from
  existing energy infrastructure.
\newblock \emph{Science}, 329\penalty0 (5997):\penalty0 1330--1333, 2010.
\newblock \doi{10.1126/science.1188566}.

\bibitem[Distefano and D'Alessandro(2023)]{Distefano2023}
Tiziano Distefano and Simone D'Alessandro.
\newblock Introduction of the carbon tax in italy: Is there room for a
  quadruple-dividend effect?
\newblock \emph{Energy Economics}, 120, 2023.
\newblock \doi{10.1016/j.eneco.2023.106578}.

\bibitem[Dowlatabadi(2000)]{Dowlatabadi2000}
H.~Dowlatabadi.
\newblock Bumping against a gas ceiling.
\newblock \emph{Climatic Change}, 46\penalty0 (3):\penalty0 391--407, 2000.
\newblock \doi{10.1023/A:1005611713386}.

\bibitem[Dubash et~al.(2022)Dubash, Mitchell, Boasson, J. Borbor-Córdova,
  Fifita, Erik Haites, Jaccard, Jotzo, Naidoo, Romero-Lankao, Shen, Shlapak,
  and Wu]{Dubash2022IPCC}
Navroz~K. Dubash, Catherine Mitchell, Elin~Lerum Boasson, Mercy
  J. Borbor-Córdova, Solomone Fifita, Erik Haites, Mark Jaccard, Frank
  Jotzo, Sasha Naidoo, Patricia Romero-Lankao, Wei Shen, Mykola Shlapak, and
  Libo Wu.
\newblock National and sub-national policies and institutions.
\newblock In P.~R. Shukla, J.~Skea, R.~Slade, A.~Al Khourdajie, R.~van Diemen,
  D.~McCollum, M.~Pathak, S.~Some, P.~Vyas, R.~Fradera, M.~Belkacemi,
  A.~Hasija, G.~Lisboa, S.~Luz, and J.~Malley, editors, \emph{Climate Change
  2022: Mitigation of Climate Change\textemdash Contribution of Working Group
  III to the Sixth Assessment Report of the Intergovernmental Panel on Climate
  Change}. Cambridge University Press, Cambridge, 2022.

\bibitem[Flora and Tankov(2023)]{Flora2023}
Maria Flora and Peter Tankov.
\newblock Green investment and asset stranding under transition scenario
  uncertainty.
\newblock \emph{Energy Economics}, 124, 2023.
\newblock \doi{10.1016/j.eneco.2023.106773}.

\bibitem[Goldberger(1986)]{Goldberger1964}
Arthur~S. Goldberger.
\newblock \emph{Econometric Theory}.
\newblock Wiley, New York, 1986.

\bibitem[Goulder(1995)]{Goulder1995}
L.H. Goulder.
\newblock Environmental taxation and the double dividend: A reader's guide.
\newblock \emph{International Tax and Public Finance}, 2\penalty0 (2):\penalty0
  157--183, 1995.
\newblock \doi{10.1007/BF00877495}.

\bibitem[Grimm et~al.(2022)Grimm, S\"{o}lch, and Z\"{o}ttl]{Grimm2022}
Veronika Grimm, Christian S\"{o}lch, and Gregor Z\"{o}ttl.
\newblock Emissions reduction in a second-best world: On the long-term effects
  of overlapping regulations.
\newblock \emph{Energy Economics}, 109, 2022.
\newblock \doi{10.1016/j.eneco.2022.105829}.

\bibitem[{IEA}(2021)]{IEA2021}
{IEA}.
\newblock Net zero by 2050\textemdash a roadmap for the global energy sector.
\newblock Technical report, International Energy Agency, Paris, 2021.

\bibitem[{IEA}(2022)]{IEA2022}
{IEA}.
\newblock World energy outlook.
\newblock Technical report, International Energy Agency, Paris, 2022.

\bibitem[Keen and Slemrod(2021)]{Keen2021}
Michael Keen and Joel Slemrod.
\newblock \emph{Rebellion, Rascals, and Revenue: Tax Follies and Wisdom Through
  the Ages}.
\newblock Princeton University Press, Princeton, 2021.

\bibitem[Kohlscheen et~al.(2021)Kohlscheen, Moessner, and
  Tak\'{a}ts]{Kohlscheen2021}
Emanuel Kohlscheen, Richhild Moessner, and Elod Tak\'{a}ts.
\newblock Effects of carbon pricing and other climate policies on
  {CO}\textsubscript{2} emissions.
\newblock Working Paper 9347, {CESifo}, 2021.
\newblock URL
  \url{https://www.cesifo.org/en/publikationen/2021/working-paper/effects-carbon-pricing-and-other-climate-policies-co2-emissions}.

\bibitem[Landis et~al.(2021)Landis, Fredriksson, and Rausch]{Landis2021}
Florian Landis, Gustav Fredriksson, and Sebastian Rausch.
\newblock Between- and within-country distributional impacts from harmonizing
  carbon prices in the {EU}.
\newblock \emph{Energy Economics}, 103:\penalty0 105585, 2021.
\newblock URL
  \url{https://www.sciencedirect.com/science/article/pii/S0140988321004540}.

\bibitem[Lee et~al.(2015)Lee, Markowitz, Howe, Ko, and Leiserowitz]{Lee2015}
Tien~Ming Lee, Ezra~M. Markowitz, Peter~D. Howe, Chia-Ying Ko, and Anthony~A.
  Leiserowitz.
\newblock Predictors of public climate change awareness and risk perception
  around the world.
\newblock \emph{Nature Climate Change}, 5\penalty0 (11):\penalty0 1014 –
  1020, 2015.
\newblock \doi{10.1038/nclimate2728}.

\bibitem[Leiserowitz(2006)]{Leiserowitz2006}
Anthony Leiserowitz.
\newblock Climate change risk perception and policy preferences: The role of
  affect, imagery, and values.
\newblock \emph{Climatic Change}, 77\penalty0 (1-2):\penalty0 45 – 72, 2006.
\newblock \doi{10.1007/s10584-006-9059-9}.

\bibitem[M\'{e}jean et~al.(2019)M\'{e}jean, Guivarch, Lef\`{e}vre, and
  Hamdi-Cherif]{Mejean2019}
Aur\'{e}lie M\'{e}jean, C\'{e}line Guivarch, Julien Lef\`{e}vre, and Meriem
  Hamdi-Cherif.
\newblock The transition in energy demand sectors to limit global warming to
  1.5\celsius.
\newblock \emph{Energy Efficiency}, 12\penalty0 (2):\penalty0 441 – 462,
  2019.
\newblock \doi{10.1007/s12053-018-9682-0}.

\bibitem[Metcalf and Stock(2020)]{Metcalf2020}
Gilbert~E Metcalf and James~H Stock.
\newblock The macroeconomic impact of {E}urope’s carbon taxes.
\newblock Working Paper 27488, National Bureau of Economic Research, 2020.

\bibitem[Nordhaus(1992)]{Nordhaus1992}
William~D. Nordhaus.
\newblock An optimal transition path for controlling greenhouse gases.
\newblock \emph{Science}, 258:\penalty0 1315--1319, 1992.

\bibitem[Ortiz-Ospina and Roser(2016)]{OWID2016}
Esteban Ortiz-Ospina and Max Roser.
\newblock Taxation.
\newblock \emph{Our World in Data}, 2016.
\newblock https://ourworldindata.org/taxation.

\bibitem[Rafaty et~al.(2020)Rafaty, Dolphin, and Pretis]{Rafaty2020}
Ryan Rafaty, Geoffroy Dolphin, and Felix Pretis.
\newblock Carbon pricing and the elasticity of {CO}\textsubscript{2} emissions.
\newblock Technical report, Energy Policy Research Group, University of
  Cambridge, 2020.
\newblock URL \url{http://www.jstor.org/stable/resrep30490}.

\bibitem[Rettig et~al.(2023)Rettig, Gärtner, and Schoen]{Rettig2023}
Leonie Rettig, Lea Gärtner, and Harald Schoen.
\newblock Facing trade-offs: The variability of public support for climate
  change policies.
\newblock \emph{Environmental Science and Policy}, 147:\penalty0 244 – 254,
  2023.
\newblock \doi{10.1016/j.envsci.2023.06.020}.

\bibitem[Riahi et~al.(2022)Riahi, Schaeffer, Arango, Calvin, Guivarch,
  Hasegawa, Jiang, Kriegler, Matthews, Peters, Rao, Robertson, Sebbit,
  Steinberger, Tavoni, and van Vuuren]{Riahi2022IPCC}
Keywan Riahi, Roberto Schaeffer, Jacobo Arango, Katherine Calvin, C\'{e}line
  Guivarch, Tomoko Hasegawa, Kejun Jiang, Elmar Kriegler, Robert Matthews, Glen
  Peters, Anand Rao, Simon Robertson, Adam~Mohammed Sebbit, Julia Steinberger,
  Massimo Tavoni, and Detlef van Vuuren.
\newblock Mitigation pathways compatible with long-term goals.
\newblock In P.~R. Shukla, J.~Skea, R.~Slade, A.~Al Khourdajie, R.~van Diemen,
  D.~McCollum, M.~Pathak, S.~Some, P.~Vyas, R.~Fradera, M.~Belkacemi,
  A.~Hasija, G.~Lisboa, S.~Luz, and J.~Malley, editors, \emph{Climate Change
  2022: Mitigation of Climate Change\textemdash Contribution of Working Group
  III to the Sixth Assessment Report of the Intergovernmental Panel on Climate
  Change}. Cambridge University Press, Cambridge, 2022.

\bibitem[Sandmo(1975)]{Sandmo1975}
Agnar Sandmo.
\newblock Optimal taxation in the presence of externalities.
\newblock \emph{The Swedish Journal of Economics}, 77\penalty0 (1):\penalty0
  86--98, 1975.
\newblock ISSN 00397318.
\newblock URL \url{http://www.jstor.org/stable/3439329}.

\bibitem[Sassi et~al.(2010)Sassi, Crassous, Hourcade, Gitz, Waisman, and
  Guivarch]{Sassi2010}
Olivier Sassi, Renaud Crassous, Jean-Charles Hourcade, Vincent Gitz, Henri
  Waisman, and C\'{e}line Guivarch.
\newblock {IMACLIM-R}: A modelling framework to simulate sustainable
  development pathways.
\newblock \emph{International Journal of Global Environmental Issues},
  10\penalty0 (1-2):\penalty0 5 – 24, 2010.
\newblock \doi{10.1504/IJGENVI.2010.030566}.

\bibitem[Seelkopf et~al.(2021)Seelkopf, Bubek, Eihmanis, Ganderson, Limberg,
  Mnaili, Zuluaga, and Genschel]{Seelkopf2021}
L.~Seelkopf, M.~Bubek, E.~Eihmanis, J.~Ganderson, J.~Limberg, Y.~Mnaili,
  P.~Zuluaga, and P.~Genschel.
\newblock The rise of modern taxation: A new comprehensive dataset of tax
  introductions worldwide.
\newblock \emph{Review of International Organizations}, 16\penalty0
  (1):\penalty0 239--263, 2021.
\newblock \doi{10.1007/s11558-019-09359-9}.

\bibitem[Semieniuk et~al.(2022)Semieniuk, Holden, Mercure, Salas, Pollitt,
  Jobson, Vercoulen, Chewpreecha, Edwards, and Viñuales]{Semieniuk2022}
G.~Semieniuk, P.B. Holden, J.-F. Mercure, P.~Salas, H.~Pollitt, K.~Jobson,
  P.~Vercoulen, U.~Chewpreecha, N.R. Edwards, and J.E. Viñuales.
\newblock Stranded fossil-fuel assets translate to major losses for investors
  in advanced economies.
\newblock \emph{Nature Climate Change}, 12\penalty0 (6):\penalty0 532--538,
  2022.
\newblock \doi{10.1038/s41558-022-01356-y}.

\bibitem[Sen and Vollebergh(2018)]{Sen2018}
Suphi Sen and Herman Vollebergh.
\newblock {The effectiveness of taxing the carbon content of energy
  consumption}.
\newblock \emph{Journal of Environmental Economics and Management}, 92\penalty0
  (C):\penalty0 74--99, 2018.
\newblock \doi{10.1016/j.jeem.2018.08.01}.

\bibitem[Stoll and Mehling(2021)]{STOLL2021}
Christian Stoll and Michael~A. Mehling.
\newblock Climate change and carbon pricing: Overcoming three dimensions of
  failure.
\newblock \emph{Energy Research \& Social Science}, 77:\penalty0 102062, 2021.
\newblock ISSN 2214-6296.
\newblock \doi{https://doi.org/10.1016/j.erss.2021.102062}.
\newblock URL
  \url{https://www.sciencedirect.com/science/article/pii/S2214629621001559}.

\bibitem[Tavoni and Tol(2010)]{Tavoni2010}
M.~Tavoni and Richard S.~J. Tol.
\newblock Counting only the hits? the risk of underestimating the costs of
  stringent climate policy: A letter.
\newblock \emph{Climatic Change}, 100\penalty0 (3):\penalty0 769--778, 2010.
\newblock \doi{10.1007/s10584-010-9867-9}.

\bibitem[Tol(1999)]{Tol1999kyoto}
Richard S.~J. Tol.
\newblock Kyoto, efficiency, and cost-effectiveness: Applications of {FUND}.
\newblock \emph{The Energy Journal}, \penalty0 (Special Issue):\penalty0
  131--156, 1999.
\newblock URL \url{https://ideas.repec.org/a/aen/journl/1999si-a06.html}.

\bibitem[Tol(2012{\natexlab{a}})]{Tol2012CCL}
Richard S.~J. Tol.
\newblock Leviathan taxes in the short run.
\newblock \emph{Climatic Change Letters}, 113\penalty0 (3-4):\penalty0
  1049--1063, 2012{\natexlab{a}}.

\bibitem[Tol(2012{\natexlab{b}})]{Tol2012EP}
Richard S.~J. Tol.
\newblock A cost-benefit analysis of the {EU} 20/20/2020 package.
\newblock \emph{Energy Policy}, 49:\penalty0 288--295, 2012{\natexlab{b}}.
\newblock \doi{10.1016/j.enpol.2012.06.018}.

\bibitem[Tol(2014)]{Tol2014en}
Richard S.~J. Tol.
\newblock Ambiguity reduction by objective model selection, with an application
  to the costs of the {EU} 2030 climate targets.
\newblock \emph{Energies}, 7\penalty0 (11):\penalty0 6886--6896, 2014.
\newblock URL \url{https://www.mdpi.com/1996-1073/7/11/6886}.

\bibitem[Tol(2023{\natexlab{a}})]{Tol2023NCC}
Richard S.~J. Tol.
\newblock Social cost of carbon estimates have increased over time.
\newblock \emph{Nature Climate Change}, 2023{\natexlab{a}}.

\bibitem[Tol(2023{\natexlab{b}})]{Tol2023bk}
Richard S.~J. Tol.
\newblock \emph{Climate economics: Economic analyses of climate, climate
  change, and climate policy (third edition)}.
\newblock Edward Elgar, Cheltenham, 2023{\natexlab{b}}.

\bibitem[Tong et~al.(2019)Tong, Zhang, Zheng, Caldeira, Shearer, Hong, Qin, and
  Davis]{Tong2019}
D.~Tong, Q.~Zhang, Y.~Zheng, K.~Caldeira, C.~Shearer, C.~Hong, Y.~Qin, and S.J.
  Davis.
\newblock Committed emissions from existing energy infrastructure jeopardize
  1.5\celsius{} climate target.
\newblock \emph{Nature}, 572\penalty0 (7769):\penalty0 373--377, 2019.
\newblock \doi{10.1038/s41586-019-1364-3}.

\bibitem[{UNEP}(2022)]{UNEP2022}
{UNEP}.
\newblock The closing window\textemdash climate crisis calls for rapid
  transformation of societies.
\newblock Emissions Gap Report 2022, United Nations Environment Programme,
  Nairobi, 2022.

\bibitem[Van~den Bremer and Van~der Ploeg(2021)]{Bremer2021}
Ton~S. Van~den Bremer and Frederick Van~der Ploeg.
\newblock The risk-adjusted carbon price.
\newblock \emph{American Economic Review}, 111\penalty0 (9):\penalty0
  2782--2810, 2021.
\newblock \doi{10.1257/aer.20180517}.

\bibitem[van~der Ploeg and Rezai(2020)]{Ploeg2020jeem}
Frederick van~der Ploeg and Armon Rezai.
\newblock The risk of policy tipping and stranded carbon assets.
\newblock \emph{Journal of Environmental Economics and Management},
  100:\penalty0 102258, 2020.
\newblock ISSN 0095-0696.
\newblock URL
  \url{https://www.sciencedirect.com/science/article/pii/S0095069618302481}.

\bibitem[van Heerden et~al.(2006)van Heerden, Gerlagh, Blignaut, Horridge,
  Hess, Mabugu, and Mabugu]{vanHeerden2006}
Jan~H. van Heerden, Reyer Gerlagh, James~N. Blignaut, Mark Horridge, S.~Hess,
  R.~Mabugu, and M.~Mabugu.
\newblock Searching for triple dividends in south africa: Fighting
  co\textsubscript{2} pollution and poverty while promoting growth.
\newblock \emph{Energy Journal}, 27\penalty0 (2):\penalty0 113--141, 2006.

\bibitem[Vandyck et~al.(2021)Vandyck, Weitzel, Wojtowicz, {Rey Los Santos},
  Maftei, and Riscado]{Vandyck2021}
Toon Vandyck, Matthias Weitzel, Krzysztof Wojtowicz, Luis {Rey Los Santos},
  Anamaria Maftei, and Sara Riscado.
\newblock Climate policy design, competitiveness and income distribution: A
  macro-micro assessment for 11 {EU} countries.
\newblock \emph{Energy Economics}, 103:\penalty0 105538, 2021.
\newblock URL
  \url{https://www.sciencedirect.com/science/article/pii/S0140988321004151}.

\bibitem[Waisman et~al.(2012)Waisman, Guivarch, Grazi, and
  Hourcade]{Waisman2012}
Henri Waisman, Céline Guivarch, Fabio Grazi, and Jean~Charles Hourcade.
\newblock The {I}maclim-{R} model: Infrastructures, technical inertia and the
  costs of low carbon futures under imperfect foresight.
\newblock \emph{Climatic Change}, 114\penalty0 (1):\penalty0 101 – 120, 2012.
\newblock \doi{10.1007/s10584-011-0387-z}.

\end{thebibliography}

\begin{table}[p]
    \centering
     \caption{Carbon tax efficacy according to 24 \textit{ex-ante} models and 5 \textit{ex-post} policy evaluations.}
    \label{tab:taxefficacy}
    \begin{tabular}{l r r r r} \hline
model & \# & mean & st.err. & prob. \\ \hline
\textsc{coffee}	&	63	&	4.883\%	&	0.584\%	& 0.000\\
\textsc{aim}	&	123	&	1.103\%	&	0.352\%	& 0.000\\
\textsc{image}	&	81	&	0.802\%	&	0.128\%	& 0.000\\
\textsc{remind}	&	286	&	0.705\%	&	0.045\%	& 0.000\\
\textsc{witch}	&	142	&	0.646\%	&	0.028\%	& 0.000\\
\textsc{gcam}	&	47	&	0.612\%	&	0.082\%	& 0.000\\
\textsc{gem-e3}	&	49	&	0.604\%	&	0.025\%	& 0.000\\
\textsc{message}	&	258	&	0.566\%	&	0.042\%	& 0.000\\
\textsc{poles}	&	134	&	0.544\%	&	0.046\%	& 0.000\\
\textsc{farm}	&	12	&	0.529\%	&	0.060\%	& 0.000\\
\textsc{prometheus}	&	6	&	0.442\%	&	0.065\%	& 0.000\\
\textsc{eppa}	&	4	&	0.373\%	&	0.040\%	& 0.000\\
\textsc{bet}	&	14	&	0.367\%	&	0.054\%	& 0.000\\
\textsc{grape}	&	17	&	0.331\%	&	0.042\%	& 0.000\\
\textsc{en-roads}	&	2	&	0.318\%	&	0.007\%	& 0.000\\
\textsc{dne21}	&	34	&	0.317\%	&	0.030\%	& 0.000\\
\textsc{muse}	&	6	&	0.238\%	&	0.099\%	& 0.000\\
\textsc{c3iam}	&	4	&	0.228\%	&	0.008\%	& 0.000\\
\textsc{tiam-ucl}	&	5	&	0.225\%	&	0.051\%	& 0.000\\
\textsc{tiam-ecn}	&	58	&	0.202\%	&	0.028\%	& 0.000\\
\textsc{gemini}	&	5	&	0.165\%	&	0.051\%	& 0.005\\
\textsc{imaclim}	&	51	&	0.121\%	&	0.021\%	& 0.995\\
\textsc{env-linkages}	&	13	&	0.005\%	&	0.006\%	& 0.000\\
\textsc{ices}	&	6	&	0.004\%	&	0.001\%	& 0.000\\ \hline
Average	&	24	&	0.597\%	&	0.194\%	& \\
Weighted average	&	24	&	0.009\%	&	0.001\%	& \\ \hline
\citet{Rafaty2020}	&		&	0.110\%	&	1.779\%	& \\
\citet{Metcalf2020}	&		&	0.125\%	&	0.013\%	& \\
\citet{Kohlscheen2021}	&		&	0.130\%	&	0.030\%	& \\
\citet{Sen2018}	&		&	0.730\%	&	0.640\%	& \\
\citet{Best2021}	&		&	2.960\%	&	0.987\%	& \\ \hline
Weighted average	&		&	0.126\%	&	0.012\%	& \\ \hline
    \end{tabular}
\caption*{\scriptsize For the selected 24 IPCC models, the table shows the number of emission reduction scenarios and the mean and its standard error of the carbon tax efficacy, that is, the carbon dioxide emission reduction in 2030 divided by the carbon tax levied in 2030. The posterior probability that the model agrees with the empirical studies in the bottom rows is in the right-most column. The table also shows the average across models and the average weighted by the inverse of the squared standard error. For the 5 empirical studies, mean and standard error of the estimated carbon efficacy are shown, as well as the weighted average across studies.}
\end{table}

\begin{table}[p]
    \centering
        \caption{Value of carbon capture and emissions as share of GDP in 2050.}
    \label{tab:taxes}
    \begin{tabular}{l r r r} \hline
model	& tax &	sequestration	&	emissions	\\
 & \$/tCO\textsubscript{2} & \%GDP & \%GDP \\ \hline
\textsc{coffee}	& 3 &	-0.07	&	0.20	\\
\textsc{aim}	& 119 &	-0.29	&	1.73	\\
\textsc{gem-e3}	& 385 &	-0.30	&	1.07	\\
\textsc{gcam}	& 1720 &	-0.31	&	-4.21	\\
\textsc{remind}	& 537 &	-1.76	&	2.66 \\
\textsc{image}	& 586 &	-2.26	&	3.35 \\
\textsc{message} & 823	&	-2.26	&	4.83	\\
\textsc{witch}	& 1204 &	-3.08	&	5.84	\\
\textsc{poles}	& 4601 &	-4.09	&	17.08	\\
\textsc{imaclim}	& 913 &	-6.56	&	7.41 \\
\textsc{grape}	& 1196 &	-7.09	&	20.58	\\
\textsc{dne21}	& 977 &	-230.08	&	301.22	\\ \hline
    \end{tabular}
\caption*{\scriptsize For the selected 10 IPCC models, the table shows the gross carbon tax revenue and the total subsidy for carbon dioxide sequestration for 2050, both as a share of Gross Domestic Product. The carbon tax is shown too.}
\end{table}

\end{document}